\documentstyle[12pt]{article}

\begin{document}

\begin{titlepage}
\vspace{1cm}

\begin{centering}

{\Large \bf Tensor Coordinates in Noncommutative  Mechanics }

\vspace{.5cm}
\vspace{1cm}

{\large Ricardo Amorim }

\vspace{0.5cm}

 Instituto de F\'{\i}sica, Universidade Federal
do Rio de Janeiro,\\
Caixa Postal 68528, 21945-970  Rio de Janeiro, Brazil\\[1.5ex]
\vspace{1cm}

\begin{abstract} A consistent classical mechanics formulation  is presented in such a way that, under quantization, it
gives a  noncommutative quantum theory with interesting new features. The Dirac formalism for constrained Hamiltonian systems is strongly used,
and  the object of noncommutativity  ${\mathbf \theta}^{ij}$ plays a fundamental rule as an independent quantity. The presented classical theory, as its quantum counterpart, is naturally invariant under the rotation group $SO(D)$.
\end{abstract}

\end{centering}

\vspace{1cm}


\vfill

\noindent{\tt amorim@if.ufrj.br}
\end{titlepage}

\pagebreak


{\bf 1.\,}Space-time noncommutativity has been a very studied subject. After the first published work\cite{Snyder}, a huge amount of papers has appeared in recent times, most of them connected with  strings\cite{Strings} and noncommutative field theories (NCFT's)\cite{NCFT}. Both   theories, which are close related
\cite{Hull}-\cite{SW}, are yet in construction, and any new contribution to the theme is welcome. 

A nice framework to study aspects on noncommutativity is given by 
the so called noncommutative quantum mechanics (NCQM), due to its simpler approach. There are several interesting works in NCQM and I cite some of them\cite{Chaichan}-\cite{Rosenbaum}. In most of these papers, the object of noncommutativity ${\mathbf \theta}^{ij}$, which essentially is the result of the commutation of two  coordinate operators, is considered as a constant matrix, although this is not the general case\cite{Snyder,Deriglazov,Carlson,Morita,DFR}. Considering ${\mathbf \theta}^{ij}$ as a constant matrix spoils the Lorentz symmetry  or correspondingly the rotation symmetry for non relativistic theories.

In a recent work \cite{Amorim1},  a version of NCQM has been presented, where not only the coordinates  ${\mathbf x}^i$ and their canonical momenta  ${\mathbf p}_i$ are  considered as operators in Hilbert space but also the objects of noncommutativity ${\mathbf \theta}^{ij}$
and their canonical conjugate momenta ${\mathbf \pi}_{ij}$.  All of these operators belong to the same algebra and have the same hierarchical level. This enlargement of the usual set of Hilbert space operators permits the theory to be invariant 
under the rotation group $SO(D)$, as showed in detail in \cite{Amorim1} . Rotation invariance, in a nonrelativistic theory, is fundamental if one  intends to describe any physical system in a consistent way. 
In NCFT's it is possible to achieve the corresponding $SO(D,1)$  invariance also by promoting ${\mathbf \theta}^{\mu\nu}$ from a constant matrix
to a tensor operator\cite{Carlson}-\cite{Saxell}, although in this last situation the rules are quite different from those found in NCQM, since in a quantum field theory  the relevant operators are not coordinates but fields.

In Ref. \cite{Amorim1}, accordingly to the discussion given above, it was introduced the canonical commutator algebra\footnote{ $i,j=1,2,...,D$; $ \mu,\nu=0,1,....,D$. Natural units are adopted, where $\hbar=c=1$ .}  

\begin{equation}
\label{1}
[{\mathbf x}^i,{\mathbf p}_j] = i \delta^i_j
\end{equation}

\begin{equation}
\label{2}
[{\mathbf \theta}^{ij},{\mathbf \pi}_{kl}] = i \delta^{ij}_{\,\,\,kl}
\end{equation} 

\noindent where $\delta^{ij}_{\,\,\,kl}=\delta^{i}_{k}\delta^{j}_{l}-\delta^{i}_{l}\delta^{j}_{k}$.  It was also assumed that

\begin{equation}
\label{3}
[{\mathbf x}^i,{\mathbf x}^j] = i {\mathbf \theta}^{ij}
\end{equation}

Expression (\ref{3}) is  fundamental  in NCQM, although in its right hand side ${\mathbf \theta}^{ij}$ is usually considered as a constant antisymmetric matrix, which obviously does not satisfies a relation like (\ref{2}).  For simplicity, in \cite{Amorim1}

\begin{equation}
\label{4}
[{\mathbf x}^i,{\mathbf \theta}^{jk}] = 0
\end{equation}

\noindent and
\begin{equation}
\label{5}
[{\mathbf \theta}^{ij},{\mathbf \theta}^{kl}] = 0
\end{equation} 

\noindent Expressions (\ref{3}-\ref{5}) are  simpler than the corresponding ones, which appear in  Snyder's  paper \cite{Snyder}. They have been proposed, in the context of quantum gravity, in   \cite{DFR}. They also appear, in the context of NCFT's, in Refs. \cite{Carlson}-\cite{Saxell}. Furthermore, it is assumed in \cite{Amorim1} that

\begin{equation}
\label{6}
[{\mathbf p}_i,{\mathbf \theta}^{jk}] = 0
\end{equation}

\noindent and

\begin{equation}
\label{7}
[{\mathbf p}_i,{\mathbf \pi}_{jk}] = 0
\end{equation}

The Jacobi identity
formed with the operators ${\mathbf x}^i$, ${\mathbf x}^j$ and ${\mathbf \pi}_{kl}$ leads to 

\begin{equation}
\label{8}
[[{\mathbf x}^i,{\mathbf \pi}_{kl}],{\mathbf x}^j]- [[{\mathbf x}^j,{\mathbf \pi}_{kl}],{\mathbf x}^i]   =   - \delta^{ij}_{\,\,\,\,kl}
\end{equation}

\noindent with solution

\begin{equation}
\label{9}
[{\mathbf x}^i,{\mathbf \pi}_{kl}]=-{i\over 2}\delta^{ij}_{\,\,\,\,kl}\,{\mathbf p}_j
\end{equation}

\noindent The algebraic structure given above shows that the  shifted coordinate operator\cite{Chaichan,Gamboa,Kokado,Kijanka,Calmet}

\begin{equation}
\label{10}
{\mathbf X}^i\equiv{\mathbf x}^i+{1\over 2}{\mathbf \theta}^{ij}{\mathbf p}_j
\end{equation}

\noindent  commutes with ${\mathbf \pi}_{kl}$, ${\mathbf \theta}^{ij}$ and ${\mathbf X}^j$.   The shifted coordinate operator plays a fundamental rule in NCQM, since it is possible to form a basis with its eigenvectors. This possibility is forbidden for the usual coordinate operator since its components satisfy nontrivial commutation relations among themselves. In Ref. \cite{Amorim1} the  algebraic structure (\ref{1}-\ref{10}) is discussed, and also it  is shown that
the generalized 
angular momentum operator 

\begin{equation}
\label{11}
{\mathbf J}^{ij}= {\mathbf X}^i{\mathbf p}^j-{\mathbf X}^j{\mathbf p}^i-{\mathbf \theta}^{il}{\mathbf \pi}_l^{\,\,j}+{\mathbf \theta}^{jl}{\mathbf \pi}_l^{\,\,i}
\end{equation}

\noindent closes in the $SO(D)$ algebra 

\begin{equation}
\label{12}
[{\mathbf J}^{ij},{\mathbf J}^{kl}]=i\delta^{il}{\mathbf J}^{kj}-i\delta^{jl}{\mathbf J}^{ki}-i\delta^{ik}{\mathbf J}^{lj}+i\delta^{jk}{\mathbf J}^{li}
\end{equation}

\noindent and generates the expected symmetry transformations when acting in all the operators in Hilbert space. This does not happen when one considers the usual angular momentum operator

\begin{equation}
\label{13}
{\mathbf l}^{ij}= {\mathbf x}^i{\mathbf p}^j-{\mathbf x}^j{\mathbf p}^i
\end{equation}

\noindent since

\begin{eqnarray}
\label{14}
[{\mathbf l}^{ij},{\mathbf l}^{kl}]&=&i\delta^{il}{\mathbf l}^{kj}-i\delta^{jl}{\mathbf l}^{ki}-i\delta^{ik}{\mathbf l}^{lj}+i\delta^{jk}{\mathbf l}^{li}\nonumber\\
&+&i{\mathbf \theta}^{il}{\mathbf p}^{k}{\mathbf p}^{j}-i{\mathbf \theta}^{jl}{\mathbf p}^{k}{\mathbf p}^{i}-i{\mathbf \theta}^{ik}{\mathbf p}^{l}
{\mathbf p}^{j}+i{\mathbf \theta}^{jk}{\mathbf p}^{l}{\mathbf p}^{i}
\end{eqnarray}

\noindent even if ${\mathbf \theta}^{ij}$ is not taken as a Hilbert space operator but just as 
a constant matrix.

{\bf 2.\,}In this letter I present a possible underlying classical theory that, under quantization, reproduce the algebraic structure displayed above.
The Dirac formalism\cite{Dirac} for constrained Hamiltonian systems is extensively used in this purpose. As it is well known, this formalism teaches us that when
a theory presents a complete set of second class constraints $\Xi^a=0,a=1,2...2N$, the Poisson brackets  $\{A,B\}$ between any two phase space quantities $A$, $B$ must be replaced by the Dirac brackets

\begin{equation}
\label{15}
\{A,B\}_D=\{A,B\}-\{A,\Xi^a\}\Delta^{-1}_{ab}\{\Xi^b,B\}
\end{equation}

\noindent in order that the evolution of the system respects the constraint surface given by   $\Xi^a=0$. In (\ref{15}) 

\begin{equation}
\label{16}
\Delta^{ab}=\{\Xi^a,\Xi^b\}
\end{equation}

\noindent is  the constraint matrix and $\Delta^{-1}_{ab}$ is its inverse. Its existence  is related to the fact that the constraints $\Xi^a$ are second class.
If that matrix were singular,  linear combinations of the $\Xi^a$ could  be first class. For the first situation, the number of effective degrees of freedom
of the theory is given by $2\cal{D}$ $-2N$, where $2\cal{D}$ is the number of phase space variables and $2N$ is the number of second class constraints.
If phase space is spanned only by the $2{\cal{D}}=2D+2{\frac{D(D-1)}{2}}$ variables $x^i, p_i,\theta^{ij}$ and $\pi_{ij}$, the introduction of second class constraints  generates an over constrained theory, when compared with the algebraic structure given in \cite{Amorim1}. So it seems necessary to enlarge phase space by $2N$ variables, and to introduce at the same time $2N$ second class constraints. The simpler way to implement these ideas without spoiling  symmetry under rotations is to enlarge phase space by a pair of canonical variables $Z^i, K_i$, introducing at the same time
a set of  second class constraints $\Psi^i,\Phi_i$. 

Considering this set of phase space variables, it follows by construction the  fundamental ( non vanishing ) Poisson bracket structure

\begin{eqnarray}
\label{17}
\{x^i,p_j\}&=&\delta^i_j
\nonumber\\
\{\theta^{ij},\pi_{kl}\}&=&\delta^{ij}_{\,\,\,\,kl}
\nonumber\\
\{Z^i,K_j\}&=&\delta^i_j
\end{eqnarray}

\noindent and the Dirac brackets structure is derived in accordance with the form of the second class constraints, subject that will be discussed in what follows.

I assume that  $Z^i$ has  dimension of length $L$ , as $x^i$. This implies that both $p_i$ and $K_i$ have dimension of $L^{-1}$.
As $\theta^{ij}$ and $\pi_{ij}$ have respectively dimensions of $L^2$ and $L^{-2}$,  the simpler form of the  constraints $\Psi^i$ and $\Phi_i$ 
is given by
$\Psi^i=Z^i+\alpha x^i+\beta\theta^{ij}p_j+\gamma\theta^{ij}K_j$ and 
$\Phi_i=K_i+\rho p_i+\sigma\pi_{ij}x^j+\lambda\pi_{ij}Z^j$,
if only adimensional parameters $\alpha, \beta, \gamma, \rho, \sigma$ and $\lambda$ are introduced and any potence higher than two in phase space variables is  discarded. I could display all of these parameters 
along the implementation of the Dirac formalism. Actually this has been done, and at the end of the calculations the parameters have been chosen in order to generate, under quantization,  the commutator structure appearing in (\ref{1}-\ref{9}). The results are surprisingly simple.  The constraints  reduce, in this situation, to

\begin{eqnarray}
\label{18}
\Psi^i&=&Z^i-{1\over2}\theta^{ij}p_j
\nonumber\\
\Phi_i&=&K_i-p_i
\end{eqnarray}

\noindent and the corresponding constraint matrix (\ref{16}) becomes

\begin{equation}
\label{19}
(\Delta^{ab})=
\pmatrix{\{\Psi^i,\Psi^j\}&\{\Psi^i,\Phi_j\}\cr
	\{\Phi_i,\Psi^j\}&\{\Phi_i,\Phi_j\}}
=
\pmatrix{0&\delta^i_{j}\cr
	-\delta^{j}_i&0}
\end{equation}

A point to be stressed here is that (\ref{19}) is regular even if $\theta^{ij}$ is singular. This guarantees that the proper commutative limit of the theory can be taken. Now the inverse of (\ref{19}) is trivially given by

\begin{equation}
\label{20}
(\Delta^{-1}_{ab})=
\pmatrix{0&-\delta^{\,\,j}_i\cr
	\delta^{i}_{\,\,j}&0}
\end{equation}

\noindent and the  Dirac brackets involving only the original set of  phase space variables is 

\begin{equation}
\label{21}
\matrix{\{x^i,p_j\}_D=\delta^i_j&\,\,\,\,\,\{x^i,x^j\}_D=\theta^{ij}\cr
        \{p_i,p_j\}_D=0& \,\,\,\,\,\{\theta^{ij},\pi_{kl}\}_D=\delta^{ij}_{\,\,\,kl}\cr
        \{\theta^{ij},\theta{kl}\}_D=0&\,\,\,\,\,\{\pi_{ij},\pi_{kl}\}_D=0\cr
        \{x^i,\theta^{kl}\}_D=0&\,\,\,\,\,\,\,\,  \{x^i,\pi_{kl}\}_D=-{1\over2}\delta^{ij}_{\,\,\,kl}p_j\cr
        \{p_i,\theta^{kl}\}_D=0&\,\,\,\,\,\{p_i,\pi_{kl}\}_D=0}
\end{equation}

\noindent which gives the desired result. Actually, if $y^A$ represents phase space variables and ${\mathbf y}^A$ the corresponding Hilbert space operators, the Dirac quantization prescription   $\{y^A,y^B\}_D\rightarrow {1\over i}[{\mathbf y}^A,{\mathbf y}^B]$ gives the commutators (\ref{1}-\ref{9}).
For completeness, the remaining Dirac brackets involving $Z^i$ and $K_i$ are  here
displayed:

\begin{equation}
\label{22}
\matrix{\{Z^i,K_j\}_D=0&\,\,\,\,\,\{Z^i,Z^j\}_D=0\cr
        \{K_i,K_j\}_D=0&\,\,\,\,\,\{Z^i,x^j\}_D=-{1\over2}\theta^{ij}\cr
        \{K_i,x^j\}_D=-\delta^i_j&\,\,\,\,\,\{Z^i,p_j\}_D=0\cr
        \{K_i,p_j\}_D=0&\,\,\,\,\,\{Z^i,\theta^{kl}\}_D=0\cr
        \{Z^i,\pi_{kl}\}_D={1\over2}\delta^{ij}_{\,\,\,kl}p_j&\,\,\,\,\,\{K^i,\theta^{kl}\}_D=0\cr
        \{K_i,\pi_{kl}\}_D=0&{}}
\end{equation}

As one can verify, the only non trivial Jacobi identities involving the Dirac brackets appearing in (\ref{21}-\ref{22}) are 
given by 

\begin{eqnarray}
\label{23}
J_1&=&\{\{x^i,\pi_{kl}\}_D,x^j\}_D+ \{\{\pi_{kl},x^j\}_D,x^i\}_D+\{\{x^j,x^i\}_D,\pi_{kl}\}_D\nonumber\\ 
J_2&=&\{\{x^i,\pi_{kl}\}_D,Z^j\}_D+ \{\{\pi_{kl},Z^j\}_D,x^i\}_D+\{\{Z^j,x^i\}_D,\pi_{kl}\}_D 
\end{eqnarray}

\noindent and both $J_1$ and $J_2$  vanish identically, as expected.
Of course, (\ref{22}) is just an auxiliary set, since due to the constraints (\ref{18}), $Z^i$ and $K_i$ can be seen as dependent variables. After using the Dirac brackets, those constraints can be used in a strong way. 

{\bf 3.\,}In this classical theory the shifted coordinate

\begin{equation}
\label{24}
X^i=x^i+{1\over2}\theta^{ij}P_j
\end{equation}

\noindent which corresponds  to the operator (\ref{10}), also plays a fundamental role. As can be verified,

\begin{equation}
\label{25}
\matrix{\{X^i,X^j\}_D=0&\,\,\,\,\,\{X^i,p_j\}_D=\delta^i_j\cr
        \{X^i,x^j\}_D={1\over2} \theta^{ij}& \,\,\,\,\,\{X^i\theta^{kl},\pi_{kl}\}_D=0\cr
        \{X^i,\pi_{kl}\}_D=0&\,\,\,\,\,\{X^i,Z^j\}_D=-{1\over2} \theta^{ij}\cr
        \{X^i,K_j\}_D=\delta^i_j&\,\,\,\,\,\,\,\,  }
\end{equation}

\noindent and so the angular momentum tensor

\begin{equation}
\label{26}
{ J}^{ij}= { X}^i{ p}^j-{ X}^j{ p}^i-{ \theta}^{il}{ \pi}_l^{\,\,j}+{ \theta}^{jl}{ \pi}_l^{\,\,i}
\end{equation}

\noindent closes in the classical $SO(D)$ algebra, by using Dirac brackets in place of commutators. Actually

\begin{equation}
\label{27}
\{{ J}^{ij},{ J}^{kl}\}_D=\delta^{il}{ J}^{kj}-\delta^{jl}{ J}^{ki}-\delta^{ik}{ J}^{lj}+\delta^{jk}{ J}^{li}
\end{equation}

As in the quantum case,  the proper symmetry transformations over all the phase space variables are generated by (\ref{26}).
By writing

\begin{equation}
\label{28}
\delta A=-{1\over2}\epsilon_{kl}\{A,{ J}^{kl}\}_D
\end{equation}

\noindent one arrives at

\begin{eqnarray}
\label{29}
\delta X^i&=&\epsilon ^i_{\,\,j} X^j\nonumber\\ 
\delta x^i&=&\epsilon ^i_{\,\,j} x^j\nonumber\\
\delta p_i&=&\epsilon _i^{\,\,j} p_j\nonumber\\
\delta \theta^{ij}&=&\epsilon ^i_{\,\,k} \theta^{kj}+ \epsilon ^j_{\,\,k} \theta^{ik}\nonumber\\
\delta \pi_{ij}&=&\epsilon _i^{\,\,k} \pi_{kj}+ \epsilon _j^{\,\,k} \pi_{ik}\nonumber\\
\delta Z^i&=&{1\over2}\epsilon ^i_{\,\,j} \theta^{jk}p_k\nonumber\\
\delta K_i&=&\epsilon _i^{\,\,\,j} p_j
\end{eqnarray}

\noindent The last two equations also give the proper result on the constraint surface. So it was possible to generate all the desired structure displayed in \cite{Amorim1} by using the Dirac brackets and the constraints (\ref{18}). These constraints, as well as the fundamental Poisson brackets (\ref{17}), can be trivially generated by the first order action

\begin{equation}
\label{30}
S=\int dt \,\,L_{FO}
\end{equation}

\noindent where

\begin{equation}
\label{31}
L_{FO}=p.\dot x+K.\dot Z+\pi.\dot\theta-\lambda_a\Xi^a-H
\end{equation}

\noindent The  $2D$ quantities $\lambda_a$ are Lagrange multipliers to implement the constraints $\Xi^a=0$ given by (\ref{18}), and $H$ is some Hamiltonian. The dots "." represent internal products. Strictly, the
momenta canonically conjugate of the Lagrange multipliers are primary constraints that, when conserved, generate the secondary constraints $\Xi^a=0$. Since these  last constraints are second class, they are automatically conserved by the theory, and the Lagrange multipliers are determined in the process.

In Ref. \cite{Amorim1}, besides the introduction of the referred algebraic structure, a specific Hamiltonian has been given, representing a generalized isotropic harmonic oscillator, which contemplates with dynamics not only   the usual vectorial coordinates but also the noncommutativity sector spanned by the tensor quantities $\theta$ and $\pi$. The corresponding classical Hamiltonian 
can be written as

\begin {equation} 
\label{32}
{ H}={\frac{1}{2m}}{ p}^2+{\frac{m\omega^2}{2}}{ X}^2+
{\frac{1}{2\Lambda}}{ \pi}^2+{\frac{\Lambda\Omega^2}{2}}{ \theta}^2
\end {equation}

\noindent  which is invariant under rotations. In (\ref{32}) $m$ is a mass, $\Lambda$ is a parameter with dimension of $L^{-3}$, and $\omega$ and $\Omega$ are frequencies.  Other choices for the Hamiltonian  can be done without spoiling the algebraic structure discussed above.

The classical system given by (\ref{30}-\ref{32}) represents two independent isotropic oscillators in $D$ and ${\frac{D(D-1)}{2}}$ dimensions, expressed in terms of variables $X^i$, $p_i$, $\theta^{ij}$ and $\pi_{ij}$. The solution is elementary, but when one expresses the oscillators in terms of physical variables $x^i$, $p_i$, $\theta^{ij}$ and $\pi_{ij}$, it arises a coupling between them, with cumbersome equations of motion. In this sense the former set of variables gives, in phase space,  the  normal coordinates that decouple both oscillators.

{\bf 4.\,}To close this letter, I comment that it was possible to generate a Dirac brackets algebraic structure that, when quantized, exactly reproduce the commutator algebra appearing in \cite{Amorim1}. The presented  theory has been proved to be invariant under the action of the rotation group $SO(D)$   and could be derived through a variational principle. 

Once this structure has been given, it is not difficult to construct a relativistic generalization of such a model. The fundamental Poisson brackets become

\begin{eqnarray}
\label{33}
\{x^\mu,p_\nu\}&=&\delta^\mu_\nu
\nonumber\\
\{\theta^{\mu\nu},\pi_{\rho\sigma}\}&=&\delta^{\mu\nu}_{\,\,\,\,\rho\sigma}\nonumber\\
\{Z^\mu,K_\nu\}&=&\delta^\mu_\nu
\end{eqnarray}

\noindent and the constraints (\ref{18}) are generalized to

\begin{eqnarray}
\label{34}
\Psi^\mu&=&Z^\mu-{1\over2}\theta^{\mu\nu}p_\nu
\nonumber\\
\Phi_\mu&=&K_\mu-p_\mu
\end{eqnarray}

\noindent  generating the invertible constraint matrix

\begin{equation}
\label{35}
(\Delta^{ab})=
\pmatrix{\{\Psi^\mu,\Psi^\nu\}&\{\Psi^\mu,\Phi^\nu\}\cr
	\{\Phi^\mu,\Psi^\nu\}&\{\Phi^\mu,\Phi^\nu\}}
=
\pmatrix{0&\eta^{\mu\nu}\cr
	-\eta^{\mu\nu}&0}
\end{equation}

\noindent The Dirac brackets between the phase space variables can also be generalized from (\ref{21}-{22}). The Hamiltonian of course can not be given by (\ref{32}), but at least for the free particle, it vanishes identically, as it uses to happen with covariant classical systems\cite{Dirac}. Also  it is necessary a new constraint, which must be first class, to generate the  reparametrization transformations. In a minimal extension of the usual commutative case, it is given by the mass shell condition $\chi=p^2+m^2=0$, but other choices are possible, giving dynamics to the noncommutativity sector or enlarging the symmetry content of the relativistic action. These ideas are under study and will be published elsewhere \cite{NEXT}.


\vskip 1cm

\begin{thebibliography}{30}
\bibitem{Snyder} H. S. Snyder, Phys. Rev. {\bf 71} (1947)  38.
\bibitem{Strings}  J. Polchinski, {\cal String Theory}, University Press, Cambridge , 1998; R. Szabo, {\cal An introduction to String Theory and D-Brane Dynamics}, Imperial College Press, London, 2004.
\bibitem{NCFT} R. J. Szabo, Phys. Repp {\bf378} (2003) 207.
\bibitem{Hull}M.R.Douglas and C. Hull, JHEP {\bf 9802} (1998) 008. 
\bibitem{Sheikh} M. M.  Sheikh-Jabbari, Phys. Lett {\bf 450} (1999) 032.
\bibitem{SW} N. Seiberg and E. Witten, JHEP {\bf 9909} (1999) 032.
\bibitem{Chaichan} M. Chaichian, M. M. Sheikh-Jabbari and A. Tureanu, Phys. Rev. Lett {\bf86} (2001) 2716.
\bibitem{Chaichan2} M. Chaichian, A. Demichec, P. Presnajder, M. M. Sheikh-Jabbari and A. Tureanu, Nucl. Phys. {\bf B 527} (2002) 149.
\bibitem{Gamboa}J. Gamboa, M. Loewe and J. C. Rojas, Phys. Rev. {\bf D 64} (2001) 067901.
\bibitem{Nair}V. P .Nair and A. P. Polychronakos, Phys. Lett {\bf B 505} (2001) 267.
\bibitem{Bellucci}Stefano Bellucci and A. Nersessian, Phys. Lett. {\bf B 542} (2002) 295. 
\bibitem{Ho}P.-M. Ho and H.-C. Kao, Phys. Rev Lett {\bf 88} (2002) 151602.
\bibitem{Espinosa}O. Espinosa and P. Gaete, "Symmetry in noncommutative quantum mechanics", het-th/0206066 (2002).
\bibitem{Deriglazov}A. A. Deriglazov, Phys. Lett. {\bf B555} (2003) 83; JHEP {\bf 303} (2003) 021.
\bibitem{Smailagic} A. Smailagic and E. Spallucci, J. Phys.{\bf A36} (2003) L467; J. Phys.{\bf A36} (2003) L517.
\bibitem{Jonke}L. Jonke and S. Meljanac, Eur. Phys. Jour. {\bf C29} (2003) 433.
\bibitem{Kokado}A. Kokado, T. Okamura and T. Saito, Phys. {\bf D 69} (2004) 125007.
\bibitem{Kijanka}A. Kijanka and P Kosinski, Phys. Rev. {\bf D 70} (2004) 127702.
\bibitem{Dadic}I. Dadic, L. Jonke and S. Meljanac, Acta Phys. Slov. {\bf 55} (2005) 145.
\bibitem{Bellucci1}S. Bellucci and A. Yeranyan, Phys. Lett. {\bf B 609} (2005) 418.
\bibitem{Calmet}X. Calmet, Phys. Rev. {\bf D 71} (2005) 085012; X. Calmet and M. Selvaggi, Phys. Rev {\bf D74} (2006) 037901.
\bibitem{Scholtz}F. G. Scholtz, B. Chakraborty, J. Govaerts and S. Vaidya, J. Phys. {\bf A 40} (2007) 14581.
\bibitem{Rosenbaum}M. Rosenbaum, J. David Vergara and L R. Juarez, Phys. Lett. {\bf A 267} (2007) 267.
\bibitem{Carlson} C. E. Carlson, C.D. Carone and N. Zobin, Phys. Rev. {\bf D 66} (2002) 075001.
\bibitem{Haghighat} M.  Haghighat and M. M. Ettefaghi, Phys. Rev {\bf D 70} (2004) 034017.
\bibitem{Carone} C. D. Carone and H. J. Kwee, Phys. Rev. {\bf D 73} (2006) 096005.
\bibitem{Ettefasghi} M. M. Ettefaghi and M. Haghighat, Phys. Rev {\bf D 75 } (2007) 125002.
\bibitem{Morita}H. Kase, K. Morita, Y. Okumura and E. Umezawa, Prog. Theor. Phys. {\bf 109} (2003) 663; K. Imai, K. Morita and Y. Okumura, Prog. Theor. Phys. {\bf 110} (2203) 989.
\bibitem{Saxell} S. Saxell {\it On general properties of Lorentz invariant formulation of noncommutative quantum field thery}, hep-th 08043341 ( 2008 ).
\bibitem{DFR} S. Doplicher, K. Fredenhagen and J. E. Roberts, Phys. Lett. {\bf B331} (1994) 29; Commun. Math. Phys. {\bf 172} (1995) 187.
\bibitem{Amorim1}R. Amorim, Tensor Operators in Noncommutative Quantum Mechanics, hep-th0804.4400.
\bibitem{Dirac}P. M. Dirac, {\cal Lectures on Quantum Mechanics}, Yeshiva University, New York, 1964; K. Sundermeyer, Constrained Dynamics, Lecture Notes in Physics 169, Springer-Verlag, Berlim, 1982; M. Henneaux and C. Teitelboim, Quantization of Gauge Systems, Princeton University Press, Princeton, 1992.
\bibitem{NEXT} R. Amorim, work in progress.
\end{thebibliography}
\end{document}